\documentclass[preprint,floats,aps,epsfig,nofootinbib,amssymb]{revtex4}
\usepackage{graphicx}
\usepackage{epsfig}
\usepackage{dcolumn}
\usepackage{bm}
\def\gsim{\lower0.5ex\hbox{$\:\buildrel >\over\sim\:$}}
\def\lsim{\lower0.5ex\hbox{$\:\buildrel <\over\sim\:$}}

\begin{document}

\preprint{
   {\vbox {
      \hbox{\bf UCI-TR-2008-12}
      \hbox{\bf  MADPH--08--1507}
      \hbox{\bf NSF-KITP-08-37}
      \hbox{\bf BNL-HET-08/6}
      }}}
\vspace*{1cm}

\title{Charged Higgs Boson Effects\\  in the Production and Decay
of a Heavy Majorana Neutrino at the LHC}

\author{Shaouly Bar-Shalom$^{a,b}$}
\email{shaouly@physics.technion.ac.il}
\author{Gad Eilam$^a$}
\email{eilam@physics.technion.ac.il}
\author{Tao Han$^{c,d}$}
\email{than@hep.wisc.edu}
\author{Amarjit Soni$^e$}%
\email{soni@bnl.gov}
\affiliation{$^a$ Physics Department, Technion-Institute of Technology, Haifa 32000, Israel\\
$^b$Department of Physics and Astronomy, University of California, Irvine, CA 92697, USA\\
$^c$Department of Physics, University of Wisconsin, Madison, WI 53706, USA\\
$^d$KITP, University of California, Santa Barbara, CA 93107, USA\\
$^e$Theory Group, Brookhaven National Laboratory, Upton, NY 11973, USA}

\date{\today}

\begin{abstract}
We consider a new  interaction  between a
heavy Majorana neutrino ($N$) and a charged Higgs boson ($H^\pm$),
and show that it  can have drastic implications on lepton number violating (LNV)
signal of same-sign dileptons at the LHC.
The  LNV signal of heavy Majorana neutrinos previously considered
at the LHC, $pp \to \ell^+ N \to \ell^+ \ell^+ W^-$,
may be overwhelmed by $pp \to \ell^+ N \to \ell^+ \ell^+ H^-$.
With the subsequent decays $H^- \to \bar t b$ or $H^- \to W^- H^0$, the heavy Majorana
neutrino production leads to the spectacular events of $\ell^+ \ell^+\ b \bar  b$ + 2 jets.
We also explore the case $m_N < m_{H^+}$,
where the decay $H^+ \to \ell^+ N$ can become the dominant
$N$-production mechanism at the LHC. In particular, we show that the process
$g b \to t H^-$ followed by $t \to b W^+$ and $H^- \to \ell^- N \to \ell^- \ell^- W^+$ could lead to another type of spectacular events of $\ell^- \ell^-\ b$ + 4 jets.
\end{abstract}

\pacs{??,??,??,??}

\maketitle

\section{Introduction}

The discovery of neutrino oscillations stands as the first direct
evidence for physics beyond the Standard Model (SM),
indicating that neutrinos are massive and that their flavors mix.
The minimal realization of beyond the SM physics that can account for the observed
sub-eV neutrino masses and mixings is constructed simply by adding heavy
right-handed neutrino fields, $N$, to the SM Lagrangian (we will denote this minimal setup by
$\nu$SM):
\begin{eqnarray}
{\cal L}_{\nu SM} \equiv {\cal L}_{SM} + {1\over 2} M N N + (Y_{H} L H N + h.c.), \label{nSMF}
\end{eqnarray}
where $M$ is the right-handed Majorana neutrino mass scale,
$L$ is an SU(2) leptonic doublet and $H$ is the SM Higgs doublet.
The $\nu$SM Lagrangian gives rise to the light neutrinos mass matrix through the
classic seesaw mechanism:
\begin{eqnarray}
m_\nu = - m_D M^{-1} m_D^T \label{seesawmass}~,
\end{eqnarray}
where $m_D=v Y_H$ and $v= \langle H \rangle \sim 175$ GeV.
Thus, $m_\nu \sim {\cal O}(10^{-2}$ eV) implies that
either $m_N \sim 10^{15}$ GeV,  $Y_H \sim {\cal O}(1)$;
or  $m_N \sim m_W$ and $Y_H \sim 10^{-7}$. Evidently, if
there are  heavy Majorana neutrinos at the  electroweak (EW) scale,
then the seesaw mechanism would seem to be somewhat unnatural and
physics beyond the classic seesaw would be needed
in order to understand the very small  Yukawa couplings associated
with the neutrino mass generation.

In this minimal $\nu$SM framework, $N$ can interact with the SM gauge bosons and Higgs
through its mixing with the light SM SU(2) $\nu_L$
fields (see e.g., \cite{book}):
\begin{eqnarray}
{\cal L}_{W} &=& -\frac{g}{\sqrt{2}} U_{\ell N} \bar \ell
\gamma^\mu P_L N W^-_\mu + h.c.  \label{YWln} ~, \\
{\cal L}_{Z} &=& -\frac{g}{2 c_W}  U_{\ell N}  \overline{\nu_\ell} \gamma^\mu
P_L N Z_\mu + h.c.  \label{YZNn} ~, \\
{\cal L}_{H} &=& -\frac{g}{2} \frac{m_N}{m_W}  U_{\ell N} \overline{\nu_\ell}
 P_R N H^0  + h.c. \label{YHNn} ~,
\end{eqnarray}
 where $P_{L,R} \equiv (1 \mp \gamma_5)/2$ and
$U_{\ell N}$ are the $\nu_{\ell,L} - N$ mixing elements.
However, with no additional assumptions or symmetries imposed on the
$\nu$SM Lagrangian in (\ref{nSMF}),
these heavy-to-light mixing elements are restricted to be vanishingly small
by the seesaw mechanism itself.
In particular, the seesaw relation in (\ref{seesawmass}) leads
to $U_{\ell N} \sim \sqrt{m_\nu/m_N}$. Therefore,
we have $U_{\ell N} \to 0$ in order to successfully
generate $m_{\nu}$ in the sub-eV range, irrespective of
whether $m_N \sim m_W$ or $m_N \sim 10^{15}$ GeV.
It follows that, within the minimal seesaw setup embedded in the $\nu$SM,
the heavy Majorana neutrinos completely decouple and no signals of $N$
are expected in collider experiments.

On the other hand, as argued above, naturalness (i.e., requiring
the neutrino Yukawa terms to be of order 1)
implies that there is new physics beyond
the minimal seesaw mechanism of the $\nu$SM type
if indeed $m_N \sim m_W$. In this case
it is, therefore, phenomenologically viable to expect that the
interactions of $N$ with the EW degrees of freedom
are not necessarily suppressed, leading to very interesting lepton-number-violating
(LNV) phenomenology mediated by $N$ at high-energy colliders
such as the Tevatron, the CERN Large Hadron Collider (LHC) and the International Linear Collider (ILC).

Indeed, collider phenomenology of  heavy Majorana neutrinos
has regained interest in the past decade due to their potential role
in generating the observed sub-eV light neutrino masses.
With the upcoming LHC and the future ILC $e^+e^-$ collider,
the search for LNV signals mediated by heavy Majorana neutrinos
is particularly well motivated. This had led to some extensive studies of heavy Majorana neutrinos
in $pp$ and $p{\bar p}$ collisions \cite{delAguila1,delAguila2,han,Wpapers,oldpp},
at an $e^+ e^-$ \cite{delAguila2,buch1}, $e^-e^-$ \cite{rizzo,ourseetohmhm}
and $e^- \gamma$ collisions
\cite{delAguila2,pilaftsis1}
and at an $e p$ machine \cite{delAguila2,buch1,ali2}.
In addition, LNV decays mediated by exchanges of heavy
Majorana neutrinos were studied in top-quark and $W$-boson decays \cite{ourtWpaper} and in rare
charged meson and lepton decays \cite{ali1}.

It should be clear that, although not always explicitly stated,
the assumption which underlies {\it all} the above studies is that at least one heavy Majorana
neutrino has an unsuppressed coupling/mixing with the SM gauge bosons
and Higgs and that $N$ production (and decays) at high-energy colliders is
induced by this coupling. That is, that $U_{\ell N} \sim {\cal O}(1)\ - $  many orders of magnitudes
larger than its naively expected size within the classic seesaw, which necessarily
implies new physics beyond the minimal $\nu$SM.

In general such new physics
can be parametrized by corrections to the $\nu$SM Lagrangian represented
as a series of effective operators ${\cal O}_i$ which are constructed using the $\nu$SM fields and whose
coefficients are suppressed by powers of $1/\Lambda$, where $\Lambda$ is the scale of
the new physics:
\begin{eqnarray}
 {\cal L} = {\cal L}_{\nu SM} + \sum_{n=5}^\infty \frac{1}{\Lambda^{n-4}} \alpha_i {\cal O}_i^n \label{eff}~.
\end{eqnarray}
For example, the dimension 6 operator \cite{our-effectiveN}:
\begin{eqnarray}
 {\cal O}_{N e \phi} = i \left(\bar N \gamma^\mu \ell_R \right) \left(\phi^T \epsilon D_\mu \phi \right)~,
\end{eqnarray}
can generate the new V+A charged interaction:
\begin{eqnarray}
{\cal L}_{W} &=& \alpha_{\ell N} \bar N \gamma^\mu P_R \ell W^+_\mu + h.c.  \label{YWlnR} ~,
\end{eqnarray}
with $\alpha_{\ell N} \sim {\cal O}(v^2/\Lambda^2)$ (i.e., when the coefficient
in front of ${\cal O}_{N e \phi}$ is naturally $\alpha_{N e \phi} \sim {\cal O}(1)$) \cite{our-effectiveN}.
Thus, if the new physics is around the TeV scale we can expect $\alpha_{\ell N} \lsim 0.1$.
However, even with this generic parametrization of new physics it is hard to see
how the unsuppressed SM-like $N \gamma^\mu P_L \ell W_\mu^+$ interaction in (\ref{YWln})
can be generated, when the new heavy Majorana fields are right-handed.
To generate such a large SM-like coupling one has to assume that other non-seesaw
or seesaw-like mechanisms exist which utilize some fine-tuned relations or extra symmetries
in the neutrino sector \cite{beyond1,beyond2,ma,beyondss,deGouvea:2006gz}

Here we will take a phenomenological approach towards the V-A coupling of
$\ell N W^+$, governed by the mixiing parameter $U_{\ell N}$,
assumed to be of ${\cal O}(1)$ a priori. EW precision data from LEP imply that
$U_{\ell N} \lsim 0.1$ if $m_N > m_Z$ \cite{kagan,LEP92} and
$U_{\ell N} \lsim 0.01$ if $m_N < m_Z$ \cite{LEP92}.
Indeed, the leading $N$ signature at the LHC is traditionally assumed
to be driven by the unsuppressed $N \ell^+ W^-$ interaction
vertex as in Eq.~(\ref{YWln}) \cite{delAguila1,delAguila2,han,Wpapers}
$p p  \to W^{+\star} \to \mu^+ N \to \mu^+ \mu^+ W^- \to \mu^+ \mu^+ j j$,
where $j$ stands for a light-quark jet.
In recent analyses \cite{han,delAguila1}, it was found
that this process may be observed at the LHC at the $5\sigma$ level,
with an integrated luminosity of 30$-$100 fb$^{-1}$,
if $U_{\ell N} \lsim 0.1$ and $m_N \lsim 200-250$ GeV.
The signal and background estimates in these studies
only apply to the specific final state $\mu^+ \mu^+ j j$ with no missing energy.
In general, however,
the $N$-production and decay patterns may be drastically altered due to additional
operators involving the interactions of $N$ with the other low-energy
degrees of freedom of the underlying new physics. This can lead to interesting new
LNV signatures which may be easier to trace.
For example,
the above Drell-Yan like process, $p p  \to \mu^+ N$,
may not be the dominant $N$-production mechanism at the LHC, in which case
new strategies for $N$-discovery should be adopted.

In this paper we wish to explore one specific example of
beyond the $\nu$SM physics, in which $N$-production
and decays may be completely altered.
Since non-zero neutrino mass necessarily requires new physics beyond the SM and since
the extension for the SM Higgs sector is well motivated in many theories
beyond the SM, it is natural to consider the interplay of both. In particular,
we will focus on a heavy Majorana neutrino potentially accessible at the LHC
with a mass in the range $10~{\rm GeV} \lsim m_N \lsim 500~{\rm GeV}$,
and the observation feasibility through its interactions with a generic new charge Higgs
boson.

The paper is organized as follows: in section II we lay out the theoretical
setup. In sections III - V we discuss $N$-phenomenology at the LHC in the presence
of the new $N-H^+$ interaction in the two cases $m_N > m_{H^+}$ and $m_N < m_{H^+}$, and
in section VI we summarize and give our concluding remarks.

\section{The Charged Higgs Boson and a Heavy Majorana Neutrino}

We first introduce a generic new $\ell N H^+$ interaction in a relative model-independent
approach
\begin{equation}
  {\cal L}_{\ell N H^+ } = \frac{g}{ \sqrt{2}} \xi_{\ell N}  \frac{m_N}{m_W} \bar N P_L \ell H^+ + h.c. \label{YHln2}~,
\end{equation}
where $\xi_{\ell N}$ are dimensionless parameters whose size
depend on the underlying new physics.

In particular,
since ${\cal L}_{\ell N H^+ }$ is a typical dimension 4 Yukawa-like term, one naturally
expects $\xi \sim {\cal O}(1)$ if the new physics contains EW-scale new scalar fields
as well as heavy Majorana neutrinos.
Alternatively, such an effective interaction can be generated
in the underlying theory by exchanges of heavy gauge-bosons or heavy fermions. In this case,
guided by the effective Lagrangian prescription
in (\ref{eff}) and by naturalness (i.e., $\alpha_i \sim {\cal O}(1)$), when these
heavy particles are integrated out we expect:
\begin{eqnarray}
 \xi \sim \frac{\sqrt{2}}{g} \frac{m_W}{m_N}\  \frac{v^2}{\Lambda^2},
\end{eqnarray}
where $\Lambda$ is roughly the mass of the new heavy particle
that gives rise to the $\ell N H^+$ interaction in (\ref{YHln2}).
Thus, when $m_N \sim m_W$ we can expect e.g., $\xi \gsim 0.1$ if $\Lambda \sim 1$ TeV.

Since there is no direct experimental constraint on $\xi$ that we know of,
we will take a phenomenological approach towards the new
$\xi_{\ell N}$, exploring the implications of ${\cal L}_{\ell N H^+}$ for
$\xi$ in the range $0 < \xi < 1$ (note that $\xi \sim {\cal O}(1)$ is also consistent with
perturbative unitarity if $m_N \lsim 700$ GeV, as was noted in \cite{ourseetohmhm}).
Then, depending on its exact size, these new $\ell N H^+$ interactions
can have surprising implications on $N$-phenomenology at high-energy colliders.
For instance, in \cite{ourseetohmhm} it was shown that such a coupling can drive a
LNV same-sign charged Higgs pair-production signal, $e^- e^- \to H^- H^-$,
at an observable rate at an ILC even if $m_N \sim 1000$ TeV.

As for collider phenomenology of N in the presence of the interactions
(\ref{YWln})-(\ref{YHNn}), (\ref{YWlnR}) and (\ref{YHln2}),
we adopt a ``one-coupling scheme" for simplicity,
assuming that only one of the mixing angles dominates,
e.g., $U_{\mu N} \gg U_{e N},\ U_{\tau N}$ and similarly, $\xi_{\mu N} \gg \xi_{e N},~\xi_{\tau N}$.
These elements will be denoted hereafter by $U \equiv U_{\mu N}$ and
$\xi \equiv \xi_{\mu N}$.

\subsection{$N$ decay in the presence of $H^\pm$}

In the framework where a new $\mu N H^+$ interaction is present,
$N$ will predominantly decay via the kinematically accessible channels
\begin{equation}
N \to W^\mp \mu^\pm ,~Z \nu_\mu,~H^0 \nu_\mu,~H^\mp \mu^\pm .
\end{equation}
The partial widths for these decay channels are given by
\begin{eqnarray}
\Gamma(N \to \mu^\pm W^\mp) &\approx&  \frac{U^2  m_N^3}{16 \pi v^2}\
 (1 + 2 r_W^{}) (1 - r_W^{})^2,\nonumber \\
\Gamma(N \to \nu_\mu Z) & \approx &  \frac{U^2  m_N^3}{16 \pi v^2}\
(1 + 2 r_Z^{}) (1 - r_Z^{})^2, \nonumber \\
\Gamma(N \to \nu_\mu H^0) & \approx &  \frac{U^2  m_N^3}{16 \pi v^2}\
 (1 - r_{H^0}^{})^2, \nonumber \\
\Gamma(N \to \mu^\pm H^\mp) & \approx &  \frac{\xi^2  m_N^3}{16 \pi v^2}\
 (1 - r_{H^+}^{})^2,
\label{partial}
\end{eqnarray}
where $r_i=m_i^2/m_N^2$.
A useful limit for illustrating
that is $m_N^2 \gg m_W^2, m_Z^2, m_{H^0}^2, m_{H^+}^2$,
in which case the total decay width of $N$ can  be conveniently written as:
\begin{equation}
 \Gamma_N \approx   \Gamma_N^0 \cdot \left(4U^2 + 2 \xi^2 \right),\quad
 \Gamma_N^0 \equiv \frac{g^2}{64 \pi} \frac{m_N^2}{m_W^2} m_N = \frac{m_N^3}{16 \pi v^2}.
 \label{width0}
\end{equation}

\begin{figure}[t]
\epsfig{file=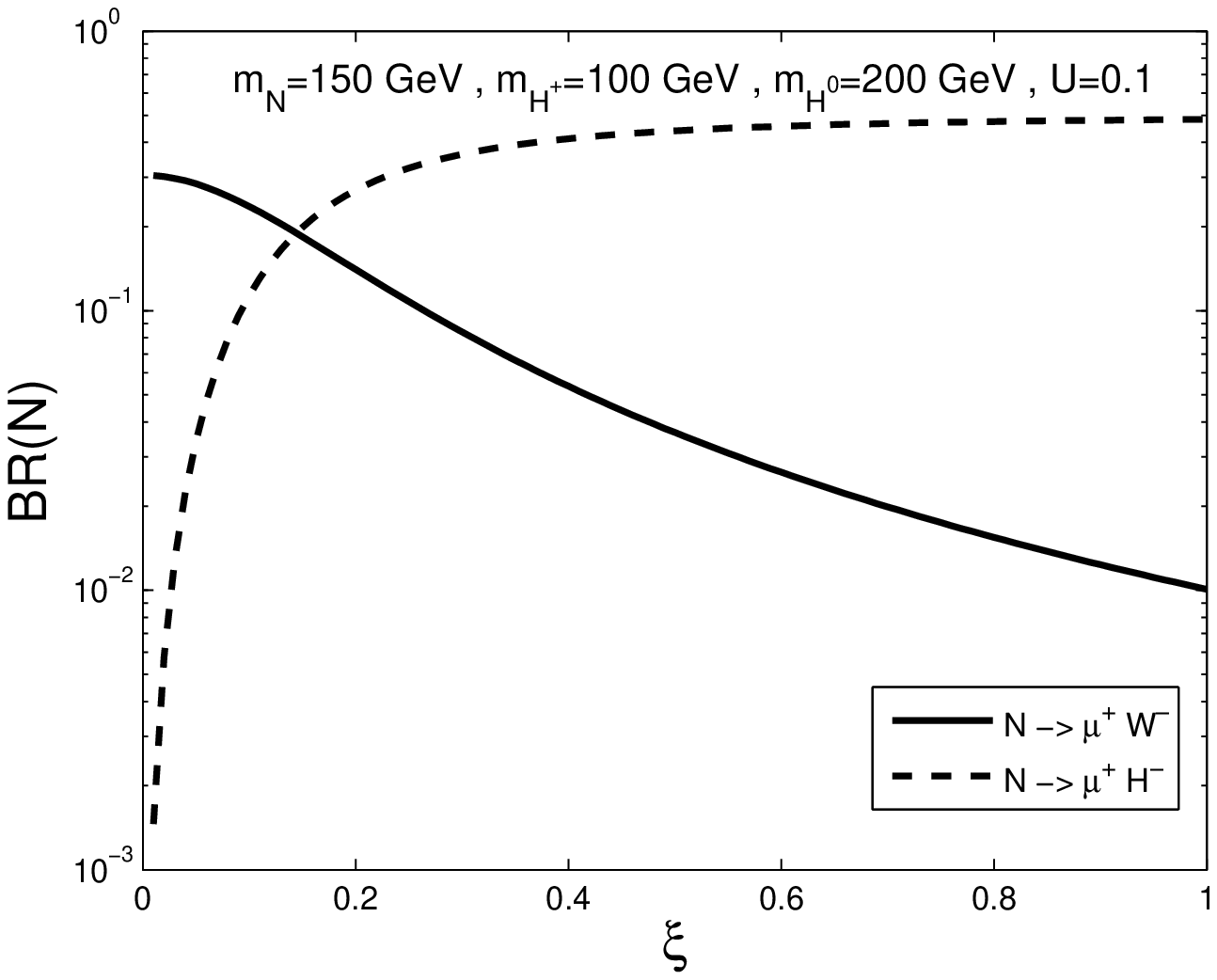,height=8cm,width=8.1cm,angle=0}
\epsfig{file=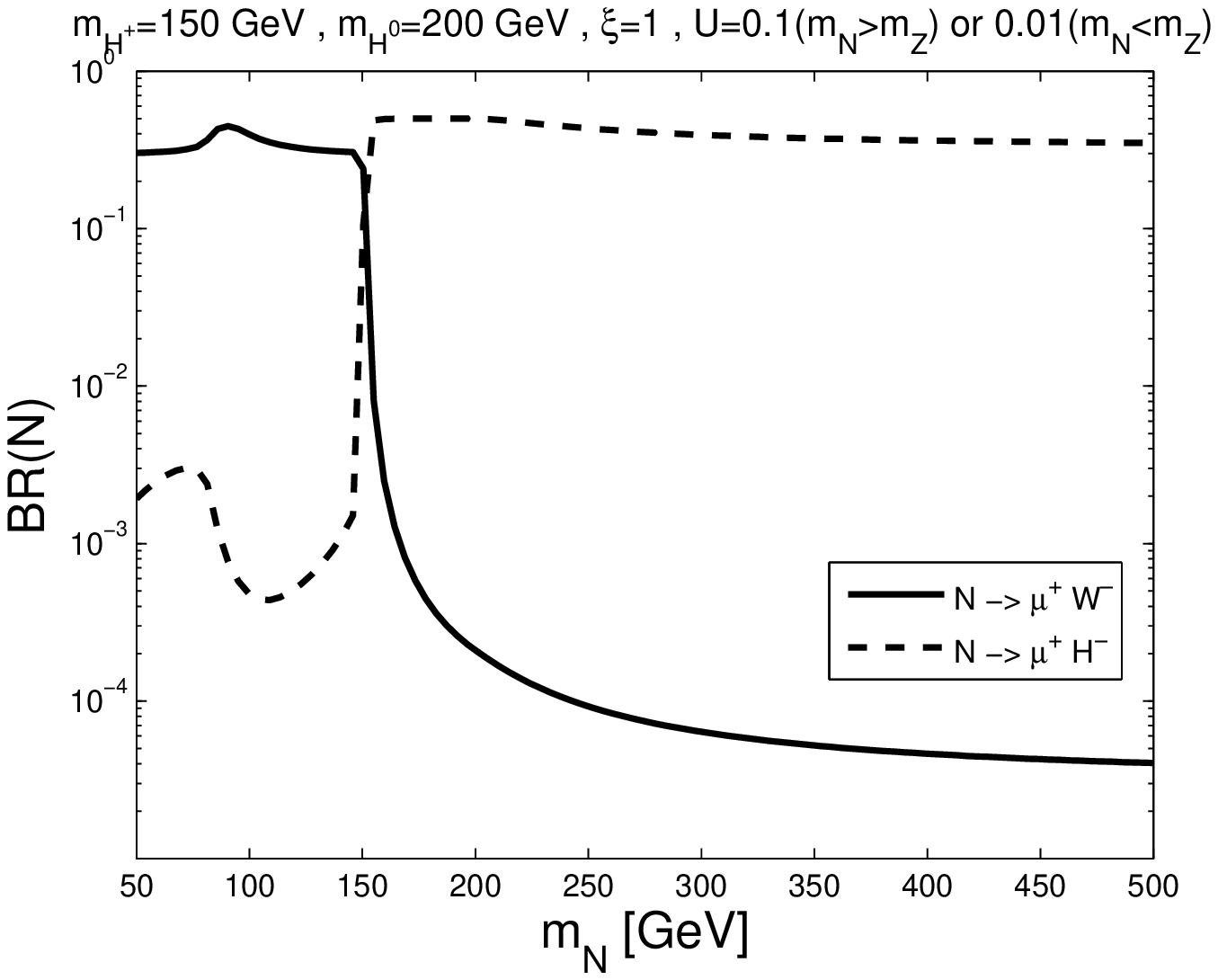,height=8cm,width=8.1cm,angle=0}
\caption{\emph{(a).
The branching fractions for the decays $N \to \mu^+ W^-$ (solid line) and
$N \to \mu^+ H^-$ (dashed line)
as a function of $\xi$, for $m_N=150$ GeV, $m_{H^+}=100$ GeV, $m_{H^0}=200$ GeV and $U=0.1$.
(b). The branching fractions for the decays $N \to \mu^+ W^-$ (solid line) and
$N \to \mu^+ H^-$ (dashed line)
as a function of $m_N$, for $m_{H^+}=150$ GeV, $m_{H^0}=200$ GeV, $\xi=1$, with
$U=0.1$ for $m_N>m_Z$ and $U=0.01$ for $m_N<m_Z$.
When $m_N < m_{H^+}$ or $m_N < m_{W}$ the curves correspond to
decays to off-shell $H^+$ or $W^+$, respectively.}}
\label{fig12}
\end{figure}

Clearly then, $N$ can become ``W/Z/$H^0$-phobic'' if $\xi^2 \gg U^2$, since in this case
its width is dominated by its decays to the new charged Higgs.
To demonstrate the effect of the $\mu N H^+$ coupling $\xi$ on the pattern of
the $N$-decays, we plot in Fig.~\ref{fig12}(a) the BR's for the decays
$N \to \mu^+ W^-$ and $N \to \mu^+ H^-$ as a function of $\xi$, for $m_N=150$ GeV
and $m_{H^+}=100$ GeV, and setting $U=0.1$ and $m_{H^0}=200$ GeV.
We can indeed see the sharp drop of the $BR(N \to \mu^+ W^-)$ as $\xi$ increases.
In particular, for $\xi \sim {\cal O}(1)$, we obtain $BR(N \to \mu^+ W^-) \sim 0.01$ while
$BR(N \to \mu^+ H^-,\ \mu^- H^+)$ saturate the decay.

In Figs.~\ref{fig12}(b) we further plot the decays
$N \to \mu^+ W^-$ and $N \to \mu^+ H^-$ as a function of $m_N$, for $\xi=1$ and for
$m_{H^+}=150$ GeV, $m_{H^0}=200$ GeV and setting $U$ to its largest allowed values
depending on $m_N$, i.e., $U=0.1$ for $m_N>m_Z$
and $U=0.01$ for $m_N<m_Z$.
Below the threshold $m_N < m_{H^+},\ m_W$, the off-shell effects have been taken
into account. Way above the threshold, the relative branchings for
$N \to \mu^+ H^-$ and $N \to \mu^+ W^-$ is governed by $\xi^2/U^2$.

\subsection{Including $H^\pm$ decays}

\begin{figure}[t]
\epsfig{file=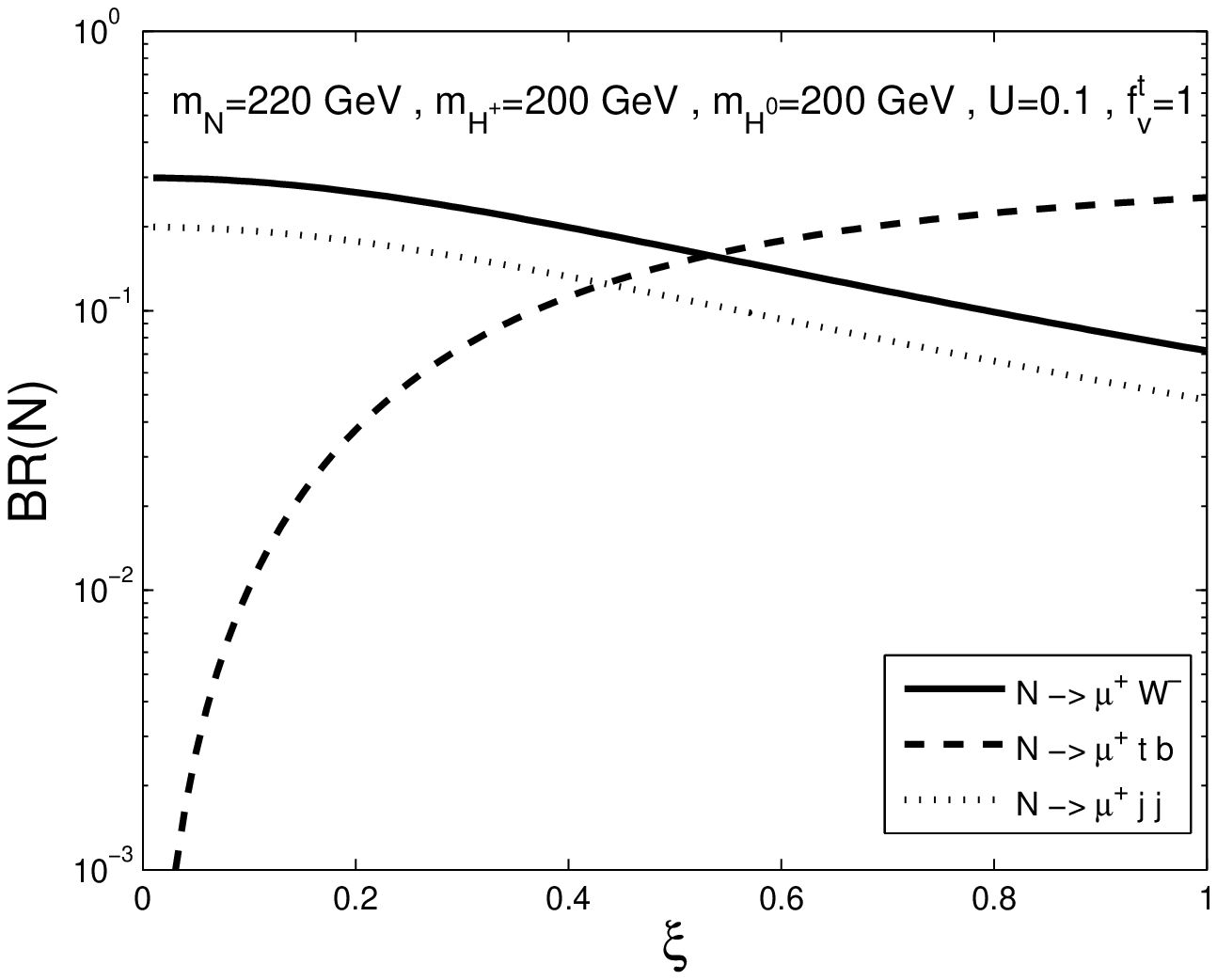,height=8cm,width=8.1cm,angle=0}
\epsfig{file=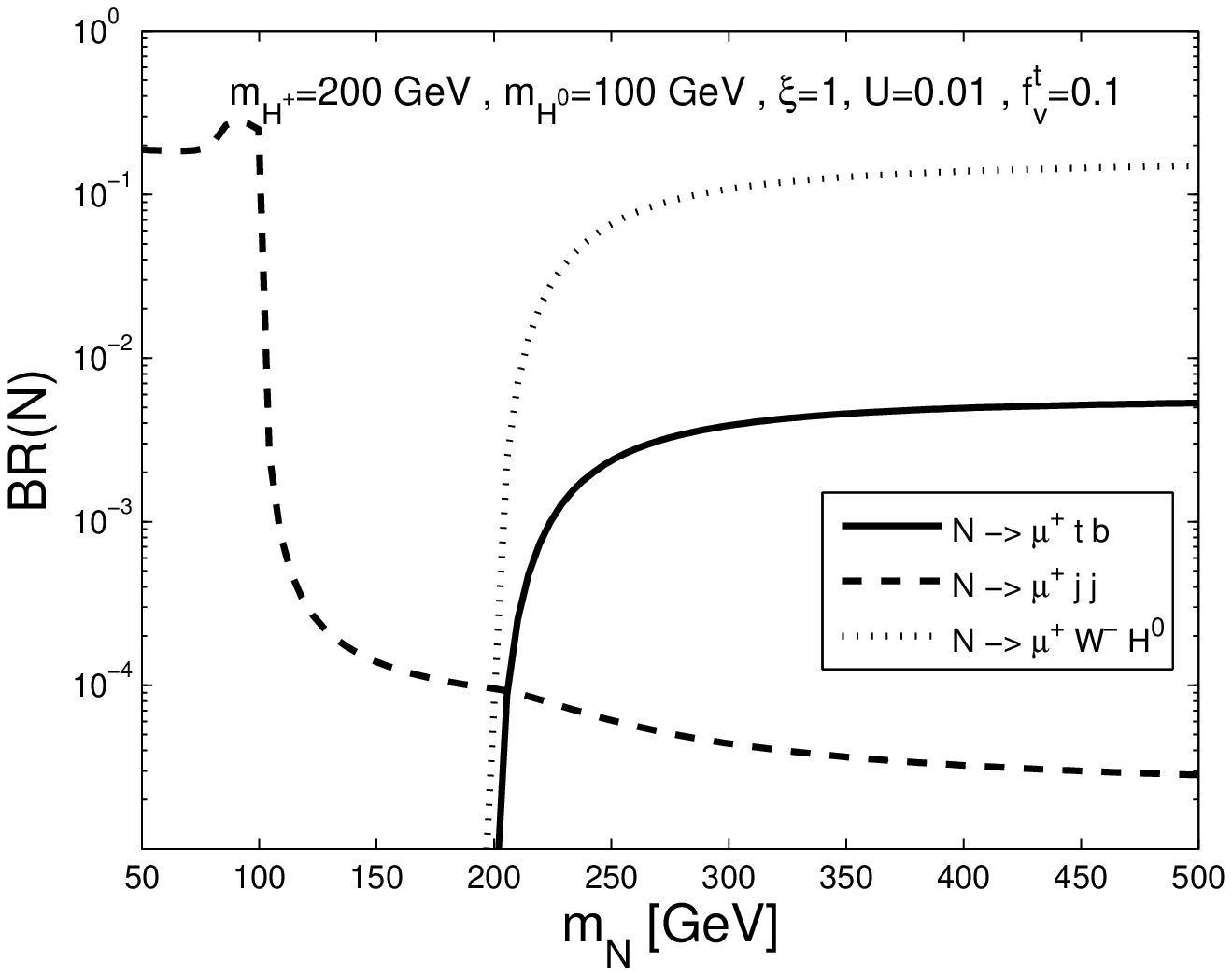,height=8cm,width=8.1cm,angle=0}
\caption{\emph{(a). The branching fractions for the decays $N \to \mu^+ W^-$ (solid line),
$N \to \mu^+ \bar t b$ (dashed line) and
$N \to \mu^+ j j$ from $W$ decay (dotted line)
as a function of $\xi$, for $m_N=220$ GeV, $m_{H^+}=200$ GeV, $m_{H^0}=200$ GeV, $U=0.1$
and $f_v^t=1$ (see text).
(b). The branching fractions for the decays  $N \to \mu^+ \bar t b$ (solid line)
$N \to \mu^+ j j$ (dashed line)  and $N \to \mu^+ W^- H^0$ (dotted line)
as a function of $m_N$, for $m_{H^+}=200$ GeV, $m_{H^0}=100$ GeV, $\xi=1$, $U=0.01$
and $f_v^t=0.1$.}}
\label{fig34}
\end{figure}

As a concrete example for illustration,
let us assume that the $H^+$-SM couplings have similar structure as
the charged Higgs couplings in generic multi Higgs doublet models:
\begin{eqnarray}
 {\cal L}_{H^+u d} &=& \frac{g}{\sqrt{2}}  \bar u \left(
f_v^u \frac{m_{u}}{m_W} P_L + f_v^d \frac{m_{d}}{m_W} P_R \right) d H^+ + h.c.  \label{YHud} ~, \\
{\cal L}_{H^+ \nu_\ell \ell} &=& \frac{g}{\sqrt{2}} f_v^\ell \frac{m_{\ell}}{m_W} \bar\nu_\ell
P_R \ell H^+ + h.c.  \label{YHln} ~, \\
{\cal L}_{H^+ H^0 W^-} &=& \frac{i g}{2} f_{W} W^+ \cdot \left( H^0 \partial H^- - \partial H^0 H^- \right) +
h.c.  \label{YHHW} ~,
\end{eqnarray}
and that the dominant $H^+$ decays channels are
\begin{equation}
H^+ \to \tau^+ \nu,~c \bar s,~t \bar b,~\mu^+ N,~W^+ H^0,
\end{equation}
when kinematically allowed.
As an example, in Fig.~\ref{fig34}(a) we show the expected BR's for the decays
$N \to \mu^+ W^-$ and $N \to \mu^+ \bar t b$ as a function of the
$\mu N H^+$ coupling strength $\xi$, setting $U=0.1$, $m_N=220$ GeV and
$m_{H^+}=m_{H^0}=200$ GeV. For the $H^+$ couplings we took
$f_v^c=f_v^t=f_v^\tau=f_{W}=1$ and $f_v^d=0$.
The channel $N \to \mu^+ j j$ from $W\to jj$ is also plotted for comparison
by the dotted curve.
We see that for this set of masses and couplings,
the channel $N \to \mu^+ \bar t b$ dominates over the previously
studied $N \to \mu^+ jj$ one when
$\xi \gsim 0.5$.

To further explore the potential deviations from ``standard''  $N$-phenomenology at future
colliders, we plot
in Fig.~\ref{fig34}(b) the BR's for the decays
$N \to \mu^+ \bar t b$,
$N \to \mu^+ j j$ and
$N \to \mu^+ W^- H^0$ for another set of inputs
and find that the $WH^0$ mode can be dominant, when $m_{H^+} > m_{H^0}$ and
if the $H^+tb$ coupling is suppressed.

In the next section we will show that these new decay channels of
N via an intermediate $H^+$, can have interesting implication for
N-phenomenology at the LHC.

\section{Same-sign di-muons signals: $pp \to \mu^+ \mu^+ W^-$ versus $pp \to \mu^+ \mu^+ H^-$}

As reiterated recently, a heavy Majorana neutrino may be searched for at the LHC
via the Drell-Yan production in \cite{delAguila1,delAguila2,han,Wpapers,oldpp}
\begin{eqnarray}
p p  \to \mu^+ N + h.c. ~,
\end{eqnarray}
with $N \to \mu^\pm W^\mp$, leading to the spectacular signal of $\mu^\pm \mu^\pm+$ 2 jets.
As discussed above, the existence of the $\ell N H^+$ interaction in (\ref{YHln2}) may
change the $N$ decay pattern significantly, possibly resulting in $N \to \mu^\pm H^\mp$ as
the dominant mode. We will study this effect in more detail below.

Denoting the cross-section $\sigma_N \equiv \sigma(pp \to \mu^+ N)$,
we can generically study the unknown couplings as parameterized in
Eqs.~(\ref{YWln}) and  (\ref{YWlnR}), and expect $\sigma_N = U^2 \sigma_{V-A} $
or $\sigma_N = \alpha_{\mu N}^2 \sigma_{V+A} $ if $U \ll \alpha_{\mu N}$, where
$U$ and $\alpha_{\mu N}$ are generic (V-A) and (V+A) $~$ $\mu N W$ couplings, respectively.
Since $\sigma_{V-A} = \sigma_{V+A}$ (up to
factors that linearly depend on the cosine of the center of mass scattering angle,
$\cos\theta$, which vanish after symmetrically integrating over phase-space),
we will generically denote
$\sigma_N \propto U^2$ regardless of whether it is generated by
the $V-A$ or the $V+A$ coupling.
In particular, for $U \sim 0.1$ (i.e., at its upper limit value), one gets
$\sigma_N \sim 100$ fb or $\sigma_N \sim 10$ fb for $m_N \sim 100$ or 200 GeV,
respectively, with no cuts \cite{delAguila1,delAguila2,han,Wpapers}.
The kinematical cuts applied to the final state particles
after the $N$ decays, e.g., to $\mu^+ \mu^+ j j$ when $N \to \mu^+ W^- \to \mu^+ j j$,
are expected to reduce these cross-sections by less than an order of magnitude
\cite{delAguila1,han}.

For $m_N$ in the range of several hundreds GeV, the total $N$ width is at most a
few percents of $m_N$. Thus, we can take the narrow width approximation (NWA):
\begin{eqnarray}
  \sigma(pp \to \mu^+ \mu^+ W^\mp) &\sim& \sigma(pp \to \mu^+ N) \times BR(N \to \mu^+ W^-) ~, \nonumber \\
  \sigma(pp \to \mu^+ \mu^+ H^\mp) &\sim& \sigma(pp \to \mu^+ N) \times BR(N \to \mu^+ H^-) ~,
 \end{eqnarray}
 as a good estimate of the total same-sign di-muon cross-sections above.
 Using the NWA, in Table~\ref{tab1} we give some sample results which compare
 between the above two signals, for the case $m_N > m_{H^+} > m_W$ and when $\xi=1$, $U=0.1$,
 without applying any kinematical cuts. To give a complete picture for the expected
 same-sign di-muons signal at the LHC, we list in Table~\ref{tab1} the total
 cross-sections summing the contributions from
 both the $\mu^+ \mu^+$ and the $\mu^- \mu^-$ channels. We denote these total
 cross-sections by
 $\sigma(pp \to \mu \mu W) \equiv \sigma(pp \to \mu^+ \mu^+ W^-) +
 \sigma(pp \to \mu^- \mu^- W^+)$ and similarly for $\sigma(pp \to \mu \mu H)$,
 reminding the reader that at the LHC the positively charged di-muons rate
is about 1.5 times larger than the negatively charged dimuons one.
As seen from Table~\ref{tab1}, the cross-section for $\mu^+ \mu^+ H^-$ can become more than
an order of magnitude larger than that of $\mu^+ \mu^+ W^-$
(similar for the $\mu^- \mu^-$ channels), if $m_N > m_{H^+}$ and $\xi \sim 1$,
even if one assumes the largest possible cross-section for $\mu^+ \mu^+ W^-$, i.e., taking $U =0.1$.
In such a case, heavy Majorana neutrinos should be searched
for at the LHC through their decay to a charged Higgs boson.
The representative cross section can be the order of $1-100$ fb, leading to $10^2-10^4$ events
with an integrated luminosity of 100 fb$^{-1}$ before acceptance cuts on the final states.

 \begin{table}[tb]
\begin{center}
\begin{tabular}{c|c|c|c|c|c|c|c}
\hline
 $m_N$ & $m_{H^+}$ & $pp \to \mu^+ N$ & $pp \to \mu^- N$& $N \to \mu^\pm W^\mp$ & $N \to \mu^\pm H^\mp$ & $pp \to \mu \mu W$ & $pp \to \mu \mu H$ \\
~[GeV] & [GeV] & $\sigma$ [fb] & $\sigma$ [fb] & BR & BR & $\sigma$ [fb] & $\sigma$ [fb] \\
\hline
~ & ~ & ~ & ~ & ~ & ~ & ~ & \\
100 & 80 & 155.6 & 106.9 & 0.008 & 0.49 & 2.1 & 128.6 \\
~ & ~ & ~ & ~ & ~ & ~ & ~ & \\
150 & 120 & 29.4 & 18.9 & 0.022 & 0.46 & 1.1 & 22.2\\
~ & ~ & ~ & ~ & ~ & ~ & ~ & \\
200 & 150 & 10.4 & 6.3 & 0.02 & 0.47 & 0.33 & 7.9 \\
~ & ~ & ~ & ~ & ~ & ~ & ~ & \\
220 & 200 & 7.4 & 4.4 & 0.07 & 0.25 & 0.83 & 2.95 \\
~ & ~ & ~ & ~ & ~ & ~ & ~ & \\
\hline
\end{tabular}
\caption{Branching fractions (BR) for the two-body decays $N \to \mu^\pm W^\mp,~\mu^\pm H^\mp$ and
cross-sections in fb for the Drell-Yan $N$-production $pp \to \mu^\pm N$ and for
the total same-sign dimuon signals at the LHC: $\sigma(pp \to \mu \mu W) \equiv \sigma(pp \to \mu^+ \mu^+ W^-) + \sigma(pp \to \mu^- \mu^- W^+)$ and similarly for $\sigma(pp \to \mu \mu H)$, see also text.
The cross-sections are evaluated using the CTEQ6M parton distribution functions
and results are given for various sets of
$N$ and $H^+$ masses, for $\xi=1$, $U=0.1$ and for $m_{H^0}=200$ GeV.
No cuts are applied.}
\label{tab1}
\end{center}
\end{table}

Of particular interest is the situation of  $H^\pm$ decay above the thresholds.
For  $m_{H^+} > m_t+m_b$,
the single-top production in association with a pair of same-sign di-muons will become
the dominant channel
 \begin{eqnarray}
  pp \to \mu^+ N \to \mu^+ \mu^+ H^- \to \mu^+ \mu^+ \bar t b  \label{mmtb}~,
 \end{eqnarray}
with a BR to be about 25\% (see Fig.~\ref{fig34}(a)) $-$
almost an order of magnitude larger than that of
$pp \to \mu^+ N \to \mu^+ \mu^+ j j$.
After the top decays, this single-top LNV channel is manifest through
$\mu^+ \mu^+ \bar t b \to \mu^+ \mu^+ b b j j$, i.e., a pair of same-sign di-muons in association
with a pair of light jets and a pair of  $b$-jets, where the two light jets reconstruct $m_W$,
which combine with one $b$-jet to reconstruct  the top mass.
 Along with the second $b$-jet, the whole $b\bar b jj$ system further reconstructs
 the charged Higgs mass.
For  $m_{H^+} > m_W+m_{H^0}$,  the new channel
\begin{eqnarray}
pp \to \mu^+ N \to \mu^+ \mu^+ H^- \to \mu^+ \mu^+ W^- H^0 ~,
\end{eqnarray}
 can become even more important, having a BR at the level of 10\% and dominating over
the $pp \to \mu^+ N \to \mu^+ \mu^+ \bar t b$ and
$pp \to \mu^+ N \to \mu^+ \mu^+ j j$ ones, if for example: $m_{H^+}=200$ GeV, $m_{H^0}=100$ GeV,
$U=0.01$, $\xi=1$ and the $H^+$ couplings:
$f_v^c=f_v^t=f_v^\tau=0.1$, $f_v^d=0$ and $f_{W}=1$ (see Fig.~\ref{fig34}(b)).\footnote{Note that
$f_{W}=1$ correspond, for example, to the coupling
in a two Higgs doublets model (e.g., in the MSSM) when $\alpha=\beta$,
where $\alpha$ and $\beta$ are the neutral Higgs mixing angles in the
CP-even and CP-odd sectors, respectively.}
Similar to the single-top LNV signal of Eq.~(\ref{mmtb}),
the LNV $\mu^+ \mu^+ W^- H^0$ channel can also induce the
signature $\mu^+ \mu^+ W^- H^0 \to \mu^+ \mu^+ b b j j$,
but with different kinematics of the final state, namely
the pair of  $b$-jets reconstruct $m_{H^0}$.

As for the signal identification, we note that there are no SM processes with $\Delta L=2$.
All backgrounds are due to some misidentification of certain sources.
Recent studies demonstrated that with judicious acceptance cuts, the SM backgrounds
to the $\mu^+ \mu^+ j j$ signal can be effectively suppressed \cite{han,delAguila1}
With the presence of two more $b$-jets coming from a top-quark decay
or from a Higgs decay, along with a pair of same-sign di-muons in the final state,
we expect to significantly improve the signal to background ratio, in comparison
with the $\mu^+ \mu^+ j j$ one. We do not plan to further quantify the signal identification
in this work.

\section{New contribution when $m_N < m_{H^+}$}
Although the spectacular decay of $N\to H^\pm \ell^\mp$ is kinematically not allowed for this case,
the inverted process $H^+ \to \ell^+ N$ can become the leading
$N$-production mechanism.
%
%
In this case the leading $H^+$ production mechanism will be
$g \bar b \to \bar t H^+$ (and similarly $g b \to t H^-$),
with a cross-section at the LHC ranging from ${\cal O}$(1000 fb) if $m_{H^+} \sim 200$ GeV to
${\cal O}$(100 fb) if $m_{H^+} \sim 600$ GeV \cite{Berger:2003sm}.
This channel has been considered as a promising channel for the discovery of a charged
Higgs boson \cite{0710.1761}.
Therefore, after $H^+$ decays via $H^+ \to \ell^+ N$, we expect the $tN$-associated
production rate at the LHC to be:
\begin{eqnarray}
 \sigma_N (g \bar b \to \bar t H^+ \to \bar t \mu^+ N) \sim (100 -1000 ~{\rm fb})
 \times {\rm BR}(H^+ \to \mu^+ N)
 \label{Nt}~,
\end{eqnarray}
  for $m_{H^+}$ of several hundreds GeV.
Then, if $\xi \sim f_v^t \sim 1$,
we expect ${\rm BR}(H^+ \to \mu^+ N) \sim
{\rm BR}(H^+ \to t \bar b) \sim {\cal O}(1)$, in which case
the $N$-signal top associated production
channel in (\ref{Nt}) can have a cross-section at the order of 100 fb, even
if $m_N \sim 500$ GeV. This is to be compared with the Drell-Yan $pp \to \mu^+ N$ production rate
which is $\sigma(pp \to W^{+\star} \to \mu^+ N) \sim 0.1$ fb for $m_N \sim 500$ GeV. Thus,
while no signal of heavy Majorana neutrinos with a mass around 500 GeV is expected at the LHC through the traditionally assumed Drell-Yan production mechanism, its production rate through $H^+$
decays can be as much as 3 orders of magnitude larger than the Drell-Yan one, possibly leading
to hundreds-thousends of heavy Majorana neutrinos
in association with a pair of $t \mu^+$ at the LHC.

When the $N$ further decays there are several interesting signatures that arise from
the above $\bar t \mu^+ N$ (or $t \mu^- N$) final state,
depending on its couplings to the SM particles.
Perhaps the simplest example is  $pp \to \mu^+ \bar t  N$ followed by $N \to \mu^+ W^-$, which will lead
to the new LNV signature
\begin{equation}
\mu^+ \mu^+ \bar t W^- \to \mu^+ \mu^+ W^- W^- b,
\end{equation}
 i.e., same sign di-muons
accompanied by an opposite sign $W$-pair and a hard $b$-jet, which reconstructs
the top-quark and the $H^+$. The hadronic decays of the $W^-W^-$ may be desirable
for the confirmation of the lepton-number violation as well as for the $m_t$ and
$m_{H^+}$ reconstruction.
A comprehensive  analysis of such new $N$-mediated LNV signal as well as
the SM backgrounds is beyond the scope of this work, although once again we do
not expect it too difficult for a signal identification.
Clearly, though, such new LNV signals deserve
to be separately addressed, since their kinematics and background
completely depart from the ones expected for the widely studied $\mu^+ \mu^+ j j$ one,
and, in particular, since hundreds to thousends of such events
are expected at the LHC if $\xi \sim 1$.

Finally, to conclude this section we note that if $m_N < m_W < m_{H^+}$, then
one might expect an enhancement in the Drell-Yan process $pp \to W^+ \to \mu^+ N$ since
the s-channel $W^+$ can resonate, giving rise to a peak in the invariant $\mu^+ N$ mass, which
might be easier to handle due to its simplicity. However, since
the limits on the $\mu W N$ coupling are more stringent in this $N$-mass range,
i.e., $U \lsim 0.01$ for $m_N \lsim m_W$ \cite{LEP92}, one expects at most (when $U \sim 0.01$)
$\sigma (pp \to W^+ \to \mu^+ N) \sim 0.1 - 10$ [fb] for
$m_N \sim 40 -80$ GeV, respectively, after applying typical LHC cuts, see e.g.,
\cite{delAguila1}. Evidently, even in the case $m_N < m_W < m_{H^+}$ we expect
$H^+ \to \mu^+ N$ to be the dominant $N$-production mechanism at the LHC.

\section{Summary and discussion}

We have argued that naturalness of the seesaw mechanism, in the sense
of having ${\cal O}(1)$ neutrino Yukawa terms, requires that either
$m_N \sim 10^{16}$ GeV or that new physics beyond the classic seesaw mechanism
exist in the neutrino sector if $m_N$ is close to the EW-scale.
Therefore, in the latter case when e.g., $m_N \sim 100 -1000$ GeV, we expect
that the new physics will generate ${\cal O}(1)$ couplings
between the heavy Majorana neutrinos and the SM particles as well as
between the heavy Majorana neutrinos and the other EW-scale degrees of freedom
of the underlying new physics. These new couplings
can then be manifest through new LNV signals, mediated by the heavy Majorana neutrinos,
at future high-energy
colliders such as the LHC and the ILC.

In this note we have considered one fairly model-independent
example of such TeV scale new physics that can
drastically change what is considered to be
the conventional $N$-phenomenology at the LHC.
In particular, we have assumed that there is
an ${\cal O}(1)$ coupling between the heavy Majorana neutrino
and a new charged Higgs of the underlying theory.
We then showed that such a new interaction term can have
interesting new implications on LNV same-sign lepton pair signals at the LHC.

For example, we found that the frequently-studied
leading LNV signal of heavy Majorana neutrinos
at the LHC, $pp \to W^+ \to \ell^+ N \to \ell^+ \ell^+ W^-$,
can become irrelevant in the presence of a sizable $\ell N H^+$ coupling,
since if ${\rm BR}(N \to \ell^+ H^-) \gg {\rm BR}(N \to \ell^+ W^-)$,
the LNV signal of a pair of same-sign charged leptons
in association with a charged Higgs, $pp \to \ell^+ N \to \ell^+ \ell^+ H^-$ become
dominant and can lead to new
LNV signatures such as $pp \to \ell^+ N \to \ell^+ \ell^+ \bar t b$ and
$pp \to \ell^+ N \to \ell^+ \ell^+ W^- H^0$.

We have also shown that, in the case $m_N < m_{H^+}$,
the decay $H^+ \to \ell^+ N$ is expected to become the dominant
$N$-production mechanism at the LHC, possibly leading to
hundreds-thousends LNV events mediated by the heavy Majorana neutrinos,
e.g., such as a new $N$-single top
associated production $pp \to \ell^- \ell^- t W^+ \to \ell^- \ell^- W^+ W^+ b$.

We did not perform any signal to background analysis, as our goal was
only to emphasize that new physics associated with heavy Majorana
neutrinos can be manifest in various, sometimes unexpected,
signals and, as such,
may call for new strategies in the search for lepton number violation
at future colliders. However, due to the additional $b$-quarks and light jets
on top of the spectacular $\mu^\pm \mu^\pm jj$ signal, we expect that the
SM backgrounds will be even easier to deal with than commonly studied
$\mu\mu jj$ channel.

\section{Acknowledgment}

SBS thanks the hospitality of the theory group in Brookhaven National
Laboratory where part of this study was performed.
The work of SBS was supported in part by NSF Grants No. PHY-0653656 (UCI),
PHY-0709742 (UCI) and by the Alfred P. Sloan Foundation.
The work of TH is supported in part by the U.S. Department of Energy under
grant No. DE-FG02-95ER40896, by the Wisconsin Alumni Research Foundation.
The work at the KITP was supported in part by the National Science Foundation
under Grant No. PHY05-51164.
The work of AS was supported in part by US DOE Contract No.
DE-AC02-98CH10886 (BNL).

\end{document}